# Enhanced broadband Terahertz radiation from two colour laser pulse interaction with thin dielectric solid target in air

Sonal Saxena[1,2], Suman Bagchi[2], Mohammad Tayyab[1,2] and Juzer Ali Chakera[1,2]
Saurabh Kumar[3] and Devki Nandan Gupta[3]

[1]Homi Bhabha National Institute, Anushakti Nagar, Mumbai 400 094, India
[2]Laser Plasma Division, Raja Ramanna Centre for Advanced Technology, Indore 452 013, India
[3]Department of Physics and Astrophysics, University of Delhi, Delhi 110007, India

E-mail: sbagchi@rrcat.gov.in

## Abstract

We report enhanced broadband Terahertz (THz) generation and detailed characterization from the interaction of femtosecond two colour laser pulses with thin transparent dielectric tape target in ambient air. The proposed source is easy to implement, exhibits excellent scalability with laser energy. Spectral characterization using Fourier transform spectrometer reveals yield enhancement of more than 150 % in the THz region of 0.1 - 10 THz with respect to conventional two-colour laser plasma source in ambient air. Further, the source spectrum extends up to 40 THz with an enhancement of flux > 30 %. Experimental results, well supported with two-dimensional particle-in-cell simulations establishes that the transient photo-current produced by the asymmetric laser pulse interaction with air plasma as well as near solid density plasma formed on the tape surface is responsible for the enhanced terahertz generation. The source will be useful for the multidisciplinary activities and ongoing applications of the laboratory-based terahertz sources.

## 1. Introduction

In recent years, a palpable surge of research in the field of terahertz (THz) frequency range is noticed owing to its unique characteristics and its potential applications in the various fields of science and technology. In the electromagnetic spectrum, THz lies in between the microwave and infrared frequencies both of which are well known to us. However, the 'THz gap' is ventured only recently revealing a new horizon to scientific community [1]. This is attributed particularly to lack of efficient sources and detectors working in this range. What makes this THz range interesting is the fact that starting from most of the solids to complicated large bio-molecules, they all have their characteristic signatures, the so called fingerprints, lying in this domain of frequencies [2]. This makes THz a very useful tool to gain insight about the internal dynamics of materials in a non-destructive, non-ionizing way unlike X-rays and intense near infrared and visible radiations. Another lucrative property of THz is that most of the common non-polar dielectrics and organic compounds are transparent to this frequency range [3].

Therefore to facilitate these scientific explorations and potential applications, a growing need for bright, broadband, table-top sources is evident [4], [5]. In this regard, two-colour laser produced plasma in ambient air medium [6] has been established as a very promising source of THz radiation compared to other commonly used laser based sources like the photoconductive antenna [7], optical rectification [8] and difference frequency generation [9] in nonlinear crystals owing to its broad THz spectral range and not limited by the source material damage as the source itself is in plasma state.

In two color laser plasma-based THz generation mechanism, femtosecond laser along with its second harmonic are co-focused in ambient air to form plasma. The air plasma creates transient current which acts as source of single-cycle pulse of THz radiation emitted in the conically forward direction [10]. The THz radiation generated from air plasma features high field intensity [11], very broadband spectral range [12] and conversion efficiency of the order of 10-4 [13]. Another remarkable benefit is the ease of implementation and no requirement of vacuum. THz radiation from this source can be focused to create electric fields of the order of MV/cm [11]. The bright, broadband THz pulse is not only useful for THz time-domain spectroscopy [14] and imaging applications [15], it can also enable various studies based on nonlinear phenomena in THz frequency range [16], [17]. As the THz radiation from this source can be generated in ambient air , laser pulses can be focused at a remote location [18] and losses due to THz beam propagation can be avoided.

However with the rapid development [19] in this field, it is only a matter of time that different research groups have focused on further development of high brightness, broadband THz sources. The major aspect of these explorations can be broadly divided into two categories, namely (1) improvisation of plasma forming medium and (2) use of high intensity lasers to create stronger plasma.

The first group of studies as mentioned above concentrates on changing the plasma medium and aiming at controlling the laser pulse propagation, thus THz generation process. A simple approach would be to change the gas species for plasma formation [20]. Another set of studies have been conducted with increase of gas pressure to enhance the electron density in plasma source [21]. Saturation in THz energy is easily observed in such experiments owing to intensity clamping in the plasma source [22]. It is well known that water molecules are strong absorbers of THz radiation. However, focusing femtosecond single color laser pulse in water placed in 5 cm long cuvette yielded THz radiation [23]. Furthermore, other liquid targets such as acetone, ethyl alcohol etc. were also used for broadband THz radiation. This counter intuitive observation was explained on the basis of in-situ generation of second harmonic radiation in the water plasma super-continuum itself. However, introduction of the two-colour laser pulses did not enhance the THz yield because of large phase dispersion in the liquid medium. Another set of studies have used continuously flowing water jet of around 170 μm thickness [24]. It was observed that a broadened chirped laser pulse of 400 fs produced larger THz energy than the minimum pulse duration. Because of these observations, it is thought that cascade ionization dominates over tunnelling ionization in the plasma formation. The THz flux has been shown to increase linearly with laser energy. In continuation to this study, addition of second harmonic has shown further enhancement in the THz flux from water jet [25]. The asymmetric field from two-colour laser has shown quadratic growth in THz flux with laser energy. Both these studies have been performed with sub-mJ laser energies. The source of THz radiation from single color laser in liquid medium is not fully understood yet. However, a major practical difficulty in these schemes lies in energy scaling of THz yield. This is due to the fact that with increase in laser energy, multiple filamentation sets in which owing to its complex behaviour disrupts the laser pulse propagation in the medium and restricts the THz production.

With the ready availability of high power 10 – 1000 TW class intense, femtosecond laser systems; formation of high temperature, solid density plasma in the laboratory at relativistic intensities is quite common these days [26]. This kind of plasma is a rich source of energetic electrons which play crucial roles in governing the expansion dynamics of the plasma manifested in emission of energetic photons and charged particles. The experiments reported in this regard are classified in two distinct categories depending on the thickness of the solid targets used.

To elucidate further, for thick solid targets (thickness >> mean free path of the average electron energy), transverse wave of fast electrons propagating along the target surface has been found responsible for dipole like emission in the THz regime [27]. On the contrary, in case of thin foil targets (thickness << mean free path of the average electron energy), the major part of the energetic electrons penetrates through the foil and escapes from the rear side of the foil. While the escape occurs, the electrons experience an abrupt change of back ground medium, solid foil to vacuum, leading to emission of transition radiation which yields broadband conical THz emission along the rear surface [28] [29]. Extensive studies exploring the role of laser and plasma (solid or foil target) conditions ably supported by extensive numerical simulations unambiguously prove this attribution. Working with thin foils have been proven to be more effective for THz yield [30].

However, all these experiments performed with such powerful lasers require special experimental conditions and are mostly aimed towards exploration of different mechanisms of THz generation. But they lack the flexibility and control to use such sources in practical applications. Apart from demanding exotic resources, these sources demand specialized operations with low repetition rate. On the other hand, use of liquid targets possesses a limitation on THz energy scaling. Evidently a lacuna exists in the development of an easily deployable, efficient, compact laboratory source which is capable of providing enhanced THz flux maintaining large bandwidth with good energy scalability. Furthermore, it should also offer ease of operation and should be devoid of any special operational requirements. In this paper, we demonstrate bright, broadband THz source based on the interaction of two-colour femtosecond laser pulses with transparent dielectric tape target kept in ambient air. The tape-based source is a simple addition to the existing schemes of THz generation in ambient air and devoid of any special requirement. Moreover, the source offers excellent scalability of THz yield with laser energy and can operate at large repetition rate, thus making it very useful for various practical applications.

## 2. Experimental Details

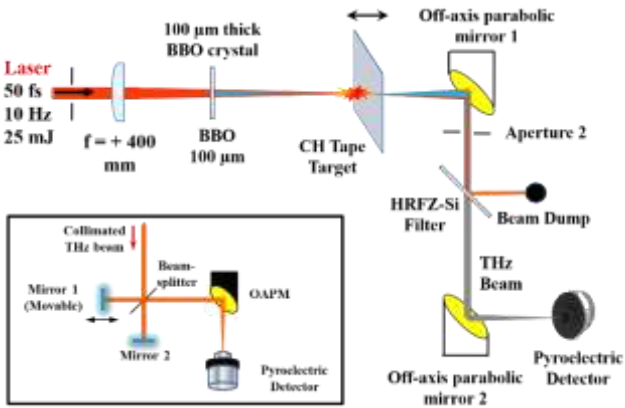

*Figure 1: Schematic diagram of the experimental setup. The inset shows the field auto-correlation set up for spectral characterization of THz radiation.*

Experiments have been performed with 50 fs, 800 nm laser pulses at 10 Hz repetition rate employing two colour laser pulses as shown in figure (1). The femtosecond laser pulses are first focussed using a + 400 mm lens in ambient air. A 100 μm thick BBO crystal is placed in the path of the converging laser beam to create co-propagating second harmonic radiation. The phase difference of the SH radiation with the fundamental laser pulse has been optimized by translating the crystal along the laser propagation direction and rotation of its axis while simultaneously monitoring the THz yield at the pyroelectric detector. A 25 μm thick transparent, dielectric, commercially available tape target made of biaxially oriented propylene (BOPP), henceforth referred as CH tape, is now placed along the path of the two-colour laser pulses in the vicinity of the plasma filament as shown in Fig.1.

Notably, choice of such a material is dictated by the fact that being transparent, dielectric and non-polar in nature, the free electron density in the CH tape is quite low and thus it allows partial transmittance of THz radiation on its own. On the contrary, any metal foil is completely opaque to THz radiation owing to its high free electron density. However, with the irradiation of the ultra-short laser pulses, the plasma formed on the surface of the CH foil conveniently exceeds the critical plasma density for THz radiation (~ $10^{16}$ cm-3). Since the emission of the THz radiation is conical [31] in shape, the formation of localized overdense plasma on the foil surface does not affect THz transmission. The CH tape target is mounted on a rotating spool which is being pulled continuously with additional adjustable clutch mechanism ensuring smooth passage of the tape without wrinkle formation. The laser interacts with a fresh position on the tape at every shot. For higher repetition operation of the source, the tape target speed can be adjusted in such a way that it always encounters with fresh surface of the tape. The tape target assembly is shown in figure (2) below. The entire assembly has been mounted on a combination of linear stages to adjust the laser irradiation on the tape surface.

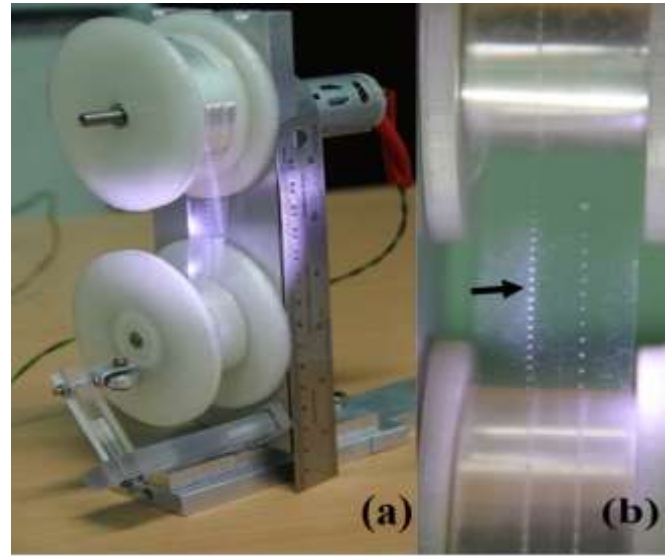

*Figure 2: (a) Tape target assembly used in experiment. (b) The arrow shows dent made by the laser on the tape.*

A 1mm thick HRFZ-Si (high resistance float zone silicon) wafer was placed in the collimated beam to function as a low pass filter and reflect the laser and other visible radiation generated from the plasma. The THz yield measurement has been performed with a pre-calibrated pyroelectric detector (THz2I-BL-BNC). For spectra characterization, an in-house developed field auto-correlation measurement system based on Fourier transform interferometry is used. Of all the conventional detection methods available for THz characterization, the Fourier transform spectrometry offers highest bandwidth support as it does not involve any non-linear property or material. In this detection method, the collimated THz beam from the first off axis parabolic mirror is divided into two parts by another HRFZ-Silicon beam splitter. Each of the beams now falls on two metallic mirrors. Upon reflection from these metallic mirrors, both the reflected beam traverses the same optical path before being focussed on to the pyroelectric detector by second off axis parabolic mirror. One of the mirrors is placed on a piezoelectric translation stage to vary the relative optical path difference between the two beams. The output signal from the pyroelectric detector is then recorded by a 16-bit data acquisition card interfaced to a computer.

## 3. RESULTS

The first step towards utilizing this tape target is optimization of THz yield. The tape target is placed initially ahead of the filament in ambient air formed by the two-colour laser pulses. In this condition, the THz yield is found to be minimal. However, as the tape is scanned across the filament length, a steady increase in THz flux is recorded. This rise in flux is limited up to a distance beyond which the flux starts reducing. So, an evident optimal condition exists here as shown in figure 3(a) for two different laser pulse energies. In continuation of the scanning when the tape has crossed the filament substantially; saturation behaviour is noticed. This is quite an intriguing behaviour which demands attention. When the tape position is ahead of the filament, i.e. towards the SH crystal (see region (1) of figure 3(b)), a mild plasma is formed only on the tape surface and there is no filament formation in air. However, as the tape gets into the filament region (region (2) of figure 3(b)) strong overdense plasma is formed on the tape surface along with plasma formation in air preceding it. Once the tape position lies within the filament, the extent of the filament is limited up to the tape surface only, as shown in figure 3(b). Once the tape is placed behind the filament (region (3) of figure 3(b)), the tape merely acts as a passive filter. Therefore, any further propagation in the forward direction does not change the THz flux resulting in apparent saturation behaviour as shown in figure 3(a). Notably, the relative position of the tape in the filament deviated considerably from the geometric focal position of the lens. When the laser pulse power increases substantially above the critical power of self-focussing, multiple filamentation starts setting in. The dynamic balance of the self-focussing and ionization induced defocussing coupled with natural beam divergence results in a small region within the filament having maximum intensity. When the tape surface coincides with this position, maximum THz yield is recorded. The tape target position is fixed here for the rest of the experiment.

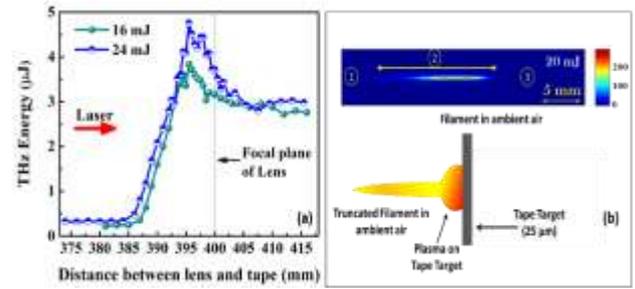

*Figure 3: (a) Variation of THz flux with scanning of the tape target along the filament. (b) Schematic of the plasma formation in ambient air without (upper) and with (lower) tape target.*

In order to compare the tape target, we first record the THz yield in ambient air with laser energy at the focussing lens. Then under identical experimental conditions, we have recorded THz yield with insertion of the tape target. The results of the THz yield variation are shown in figure (4). We have also plotted the relative "%" increase of THz flux with the incident laser energy on the same. It is evident that starting from the lowest laser energy and till the maximum laser energy, tape target consistently outperforms air plasma yield by more than 60 % increase in flux. Moreover, with increase in laser energy the relative increase in flux keeps on increasing monotonically to 90%, unlike saturation behaviour displayed by the air plasma.

It is imperative to note here that as shown in figure 3(b) earlier, in all cases the filament does not propagate beyond the tape target. Therefore, the extra yield in THz flux is indeed provided by the tape target alone. It is also important to note that the focal spot diameter measured at low laser intensity (no plasma formation) has been measured to be 20 μm (1/e2). However, as the laser power crosses the critical power substantially, the process of multiple filamentation sets in. This makes accurate measurement of filament diameter difficult. However transverse time integrated imaging of filaments with CCD camera reveals that the average filament diameter definitely exceeds an order of magnitude from the focal spot dimension recorded [32]. Moreover, with increase of laser energy; the multiple filamentation sets in early and the filament propagates with larger diameter. Consequently, increase of laser energy does not effectively increase the laser pulse intensity during propagation and hence, availability of electrons via ionization of the medium. Moreover, with increasing propagation distance in the filament, the phase slippage between the two colours (ω and 2ω) of the femtosecond laser pulse also increases. The cumulating effect of these two factors restricts the THz generation leading to saturation behaviour. On the contrary, in case of the tape target which provides high density (near solid density) bound electrons in a small distance (less than its thickness) no such saturation behaviour occurs.

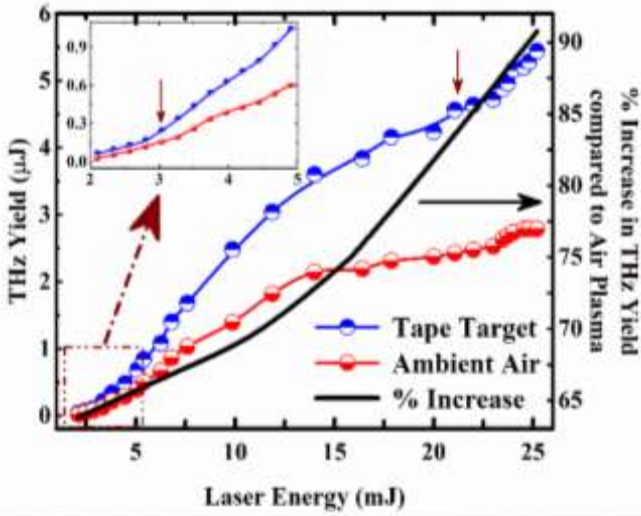

*Figure 4: Laser energy scaling of the THz flux produced by tape target in comparison with ambient air plasma. The relative "%" increase in THz yield with respect to ambient air plasma is shown on the right axis. The inset shows expanded view of the THz fluxes at low laser energies.*

The release of electrons from ionization of higher charged states in a near solid density plasma keeps contributing to plasma current density and thus (contrary to air plasma) steady increase of THz yield is observed without showing saturation behaviour on THz flux. A step like behaviour observed around laser energies of 3 mJ and 22.5 mJ as shown (downward arrows) in Fig. 4 in case of THz yield from tape target reconfirms our understanding as discussed above. At the lower laser energy case, the release of electrons appears to be contributed by $N_2$ (3+, 47.45 eV), C (3+, 47.88 eV) and $O_2$ (3+, 54.93 eV) together while for the higher energy case, it is predominantly C (4+, 64.49 eV) which provides the electrons contributing to THz generation as also confirmed by simulations described later.

Now we focus on the other important aspect of this study, spectral characterization of THz radiation produced by tape target. The spectral characterization has been performed using the in-house developed field auto-correlator (FAC) based on the principle of Fourier transform spectrometry [33] [34] adapted in THz domain (FT-THz). Contrary to conventional coherent THz detection methods (e.g. electro-optic sampling (EOS) [35] [36], air biased coherent detection (ABCD) [37,38], photo-conductive antenna (PCA) [39,40]), FT-THz is incoherent detection technique in nature. The measurement can reveal only the spectral amplitude of the THz radiation but cannot retrieve the phase information explicitly. However, the FT-THz is most suited for broadband THz detection as it is not limited by presence of any finite response (PCA), phonon resonances (EOS) or non-linearity (ABCD) of the medium during the detection process [41]. The comparative spectrum of THz emitted by air plasma with and without the presence of tape target is displayed in figure (5) below. Figures 5(a, b) display the Fourier transformed spectral domain THz signatures from both the sources in logarithmic (5(a)) and linear (5(b)) scale respectively to highlight the lower and higher frequency contents of the tape target compared to ambient air. The displayed spectrum has been averaged over 32 independent measurements in both cases. Estimating the integrated THz spectral flux (based on area under the curve) from figure 5(a or b) within a fixed frequency band of 10 THz reveals that the tape target produces more than 150 % higher yield compared to only air plasma in 0.1 – 10 THz spectral range, 60 % in 10 – 20 THz, and 30 % in 20 – 40 THz (fig. 5(c)). Similar behaviour is observed for other laser pulse energies also.

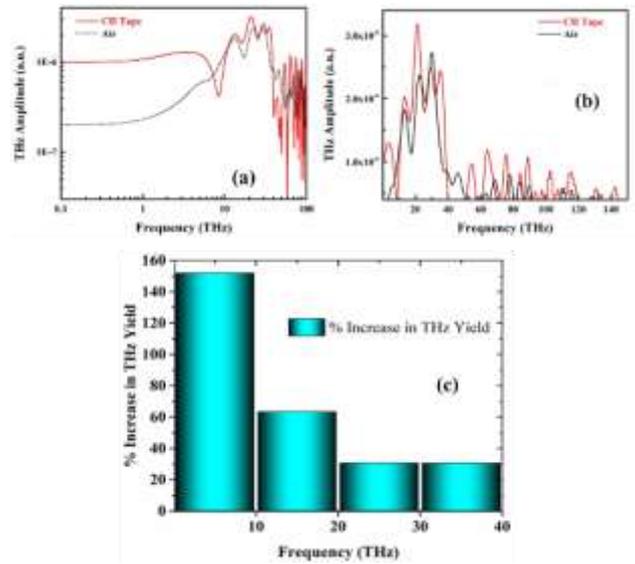

*Figure 5: Spectral characterization of tape target in ambient air and air plasma only. (a)The frequency axis is sown in logarithmic scale and (b) in linear sale to highlight the low and high frequency extent of the THz radiation respectively. (c) Frequency integrated enhancement in THz yield from tape target with respect to air plasma.*

Having described the essential properties of the new source, now we focus on the mechanisms of THz generation. The transient photo-current model [42] is most dominant and well established mechanism for THz generation from air plasma source when irradiated by the two-colour femtosecond laser pulses with a relative phase difference of "π/2". The model predicts that the resultant THz flux is proportional to the instantaneous variation of electron density in the medium. So, shortening the laser pulse duration or alternatively, rapid increase of electron density in the medium may work well to reveal an enhancement. The presented scheme relies on the second aspect of these possibilities. This behaviour is manifested in the figure (6) shown below representing THz flux variation from both, the tape target and ambient air medium with rotation angle of the second harmonic generating crystal. The variation of THz flux in ambient air only is well reproduced as observed in other experiments. However, with the addition of tape target though, the overall shape of the angle dependent flux distribution remains similar but an overall increase is noticed in almost all the directions. Moreover, the sharp features of the flux distribution as observed for air are mostly smeared out. A possibility arises that the near solid density plasma itself can be a source of second harmonic radiation which coupled with the fundamental laser wavelength can yield THz radiation.

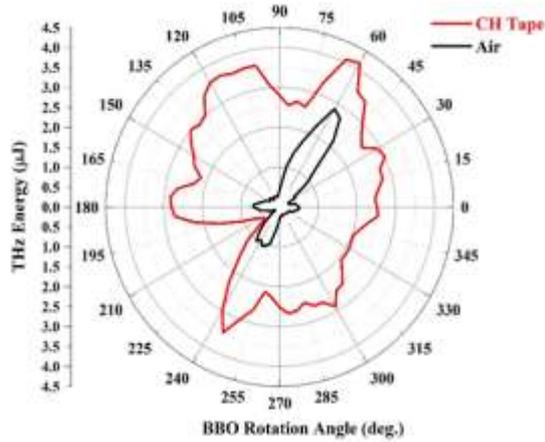

*Figure 6: Variation of Terahertz flux with rotation angle of the second harmonic generating crystal in ambient air and with tape target.*

An extended pre-plasma formed in front of the tape target surface may serve the purpose. In order to verify this, we have used single colour (only fundamental) laser pulse (with pre-pulse contrast better than 10$^4$) irradiating on the tape surface. However, in this case we have not observed any THz radiation at the detector. Moreover, the tape being transparent to the fundamental and second harmonic laser wavelengths also goes against the dominant role of pre-plasma in the present scenario. This is further supported by the observation that with two colour laser pulses, intentional increase of pre-pulse level leads to clear decrease in THz flux in the detector. The only possibility appears to be rapid phase change between the fundamental and the second harmonic laser pulse owing to the presence of near solid density plasma formed on the tape surface. In case of air plasma, the phase difference between fundamental ($\omega$) and the second harmonic ($2\omega$) radiation can be express as $\theta = (n_\omega - n_{2\omega}) d/c_0 + \varphi$, where $n_{\omega/2\omega}$ represents the refractive indices of the two colours, '$c_0$' is the speed of light in vacuum and 'd' is the distance between second harmonic crystal and the plasma. '$\varphi$' is the initial phase difference between the two beams just after crossing the second harmonic crystal. In our case, the thickness of the second harmonic crystal is barely 100 μm and thus the resultant phase mismatch '$\varphi$' can be ignored. Whereas in case of the tape target in ambient air medium, the phase matching condition is also affected by the small amount near solid density plasma. The relative phase slippage [43] between the two colour laser pulses can be expressed as: $\delta\theta = (3\pi/4) \cdot (l/\lambda) \cdot (N_e / N_{cr})$, where 'l' is the scale length of the near solid density plasma, '$\lambda$' is the wavelength corresponding to $\omega$, '$N_{cr}$' is the critical plasma density for wavelength '$\lambda$' (800 nm) and '$N_e$' is the plasma electron density. During propagation in air, considering typical filament length of around 5 mm, the corresponding maximum phase matching angle comes out to be 21°. However, when reaching solid density plasma, the extent of the pulse propagation is limited up to the critical density of $\omega$ only. Considering typical extent of the pre-plasma up to 20$\lambda$ in length (very sharp density gradient) and a step like average density profile (instead of conventional linear or exponential density profiles) with about 1% of the critical density, the relative phase slippage between the two pulses ($\omega$ and $2\omega$) yields a huge angle of 270°. This large phase slippage (albeit based on very simplified assumptions) implies evidently that the sharp BBO angle dependence of THz generation in case of air plasma is now definitely being smoothened out by the marginal extent of near solid density plasma which the laser pulses can penetrate into.

## 4. Particle-in-Cell Simulations

We have performed two dimensional particle-in-cell simulation (2D PIC) using the code EPOCH [44] to understand the THz generation from the tape target. The simulation box is 200 μm × 200 μm in dimension in x-y direction (with x being the laser propagation direction) with grid spacing of 20 nm having 5 particles per cell and is filled with nitrogen gas of density $4 \times 10^{18}$ cm$^{-3}$. The two-colour, linearly polarized along z-direction laser pulse enters the simulation box from the left side. The intensity of the fundamental and the second harmonic laser pulses are intentionally chosen to be higher ($5\times10^{16}$ W/cm$^2$ and $1\times10^{16}$ W/cm$^2$ respectively) than the experimental conditions in order to reveal role of "hot" electrons from the tape target. The tape target is simulated by introducing layers of carbon and hydrogen having densities of $4\times10^{22}$ cm$^{-3}$ and thickness 25 μm. Notably the chosen laser intensities are strong enough to ionize neutral nitrogen atom to 5+ charged states. Therefore, choice of accurate molecular structure of tape target is not a prerequisite. In order to identify the underlying mechanism, the simulations have been performed in three stages, (i) THz generation in ambient air only (ii) THz generation in tape target placed in vacuum (laser pulse propagation effects in air are therefore eliminated) and (iii) tape target placed in ambient air. Notably use of single colour laser pulse as well as two colour laser pulses polarized in y-direction does not yield any THz radiation as in experiments. This completely negates the role of ponderomotive force for THz generation. The results of 2D PIC simulations are shown in Figure (7) below. The dashed vertical lines represent the position of the tape target.

Notably particle-in-cell simulations [45,46] do not include non-linear effects into account. Only the possible role of plasma can be explored here. Because of the orders of magnitude larger electric field of two-colour laser pulses along with the local electric fields of the plasma, it is difficult to observe the THz electric field in simulation results. In order to visualize the THz electric field, the electric field magnitudes therefore have been curtailed within the limit of ±1 GV/m as shown in figure (7) below. The values of the electric fields beyond this limit have been artificially reduced to zero. However, still the noise from the plasma electric field is clearly visible.

While Figure 7(a) shows the THz generation in ambient air, Figure 7(b) displays the THz generation when the two-colour laser pulses are incident on the tape target in vacuum. In this case, we do observe relatively mild THz generation, both in the forward (laser propagation direction) and the reverse (towards laser) direction. The THz emission in the reverse is quite weak compared to forward direction indicating that the plasma itself may not be an efficient emitter of THz radiation. However, in the forward direction presence of THz radiation may indicate role of expanding electrons from the plasma. The tape target in ambient air represented in figure 7(c); shows distinct and enhanced THz flux in the forward direction. As per the transient photo-current model, the rapid change of electron density with time gives rise to THz radiation.

Following the same hypothesis, the tape target itself should be an efficient THz emitter. However, it is only in the presence of the air plasma that the THz yield from the tape target encompasses the previous two scenarios (figure 7(a) and 7(b)). We believe that the filamentary propagation of the laser pulse in ambient air at such large laser intensities invariably gives rise to multiple filamentation [47] in the laser beam. This results in formation of "hot spots" in the laser beam profile acting as local sources having higher laser intensities. Therefore, interaction of this heavily filamented beam with the tape target (larger focal spot area having sufficient intensities to cause ionization) produces a cumulative larger electron density as compared to the scenario when the tape is placed in vacuum. However, due to the presence of the plasma electric fields it has not been clearly represented in the simulation. Increase of laser energy will only increase the multiple filamentation in the beam and thus will keep on increasing the THz yield as seen experimentally in figure 4.

*Figure 7: Terahertz electric field simulated with two-dimensional particle-in-cell simulations for (a) air plasma, (b) tape target in vacuum and (c) tape target in air. (d) Terahertz spectra derived from the time evolution for the three cases. The colorbar limits have been kept the same for all three for visual comparison.*

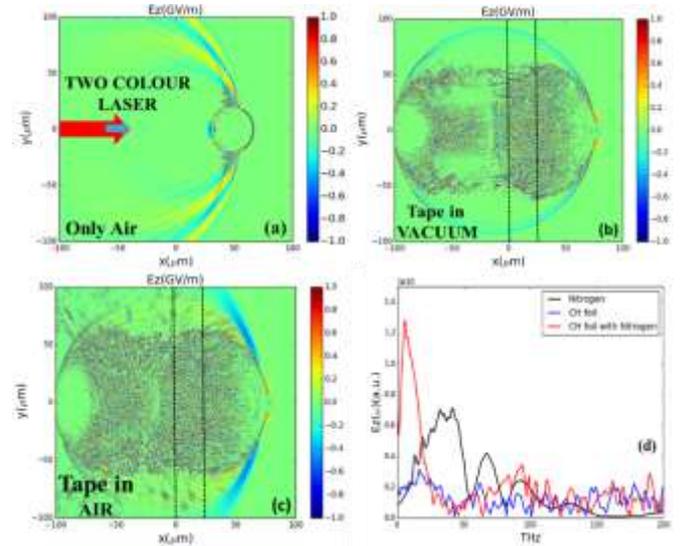

The spectra of the THz radiation (figure 7(d)) generated from the three scenarios have been derived by collecting the electric field amplitudes for the entire time evolution at a point away from the central axis at (50 µm, -50 µm) coordinate and then performing Fourier transform to convert them to spectral domain. From figure 7(d) it is evident that compared to ambient air plasma as well as from the tape target in vacuum, the combination of tape target in ambient air yields higher THz flux. However, the spectral content is not fully recovered unlike in experiments as shown in figure 6(b) earlier. This may be attributed to the limitation in computational resources as the simulations had to be truncated on a relatively shorter time scale. However, even with the existing limitations, comparing figures 7(b and c), it is clearly identifiable that unlike air plasma, the combination of tape target in air produces larger yield of THz radiation and exhibits two distinct sources of THz radiation in line with the broadened peak observed in figure 6(a). A weak component of THz radiation is also observed to be emitted in the backward direction which has not been measured in experiments.

## 5. Conclusion

In conclusion, we demonstrate enhanced broadband THz emission from the two-color femtosecond laser pulses with transparent, dielectric tape target in ambient air. The tape target shows excellent scalability with laser energy without limiting on the bandwidth and can operate at large repetition rates once the initial optimization of it position is taken care of. The tape target yields 150 % (in 0.1 – 10 THz spectral range), 60 % (in 10 – 20 THz) and 30 % (in 20 – 40 THz) higher compared to ambient air in identical experimental conditions. The experimental observations supported with 2D-PIC simulations reveal the role of transient photo-current process as being responsible for the enhancement of the THz yield. We believe with the present momentum towards THz based science and technologies worldwide, this source can be very useful in designing table top efficient THz systems requiring large flux such as in time resolved spectroscopy, explosive detection [48,49], imaging [50,51], chemistry, materials science [52,53], medical applications and many interdisciplinary areas of explorations [54], [55].

## Acknowledgements

The authors (SS, SB, MT and JAC) thank R. A. Joshi for laser operations and S. Sebastin, L. Kisku, K. C. Parmar for providing ground support during the experimental campaign. Further, the authors (SK and DNG) thank the Department of Physics and Astrophysics, University of Delhi, for providing the facility of high-performance cluster machine to carry out the simulation work.

## References

[1] G. Davies and E. Linfield, *Bridging the Terahertz Gap*, Phys. World **17**, 37 (2004).
[2] L. Ho, M. Pepper, and P. Taday, *Terahertz Spectroscopy: Signatures and Fingerprints*, Nat. Photonics **2**, 541 (2008).
[3] A. Podzorov and G. Gallot, *Low-Loss Polymers for Terahertz Applications*, Appl. Opt. **47**, 3254 (2008).
[4] H. A. Hafez, X. Chai, A. Ibrahim, S. Mondal, D. Ferachou, X. Ropagnol, and T. Ozaki, *Intense Terahertz Radiation and Their Applications*, Journal of Optics (United Kingdom).
[5] K. Reimann, *Table-Top Sources of Ultrashort THz Pulses*, Reports Prog. Phys. **70**, 1597 (2007).
[6] J. Dai, B. Clough, I. C. Ho, X. Lu, J. Liu, and X. C. Zhang, *Recent Progresses in Terahertz Wave Air Photonics*, IEEE Trans. Terahertz Sci. Technol. **1**, 274 (2011).
[7] N. M. Burford and M. O. El-Shenawee, *Review of Terahertz Photoconductive Antenna Technology*, Opt. Eng. **56**, 010901 (2017).
[8] A. Rice, Y. Jin, X. F. Ma, X. C. Zhang, D. Bliss, J. Larkin, and M. Alexander, *Terahertz Optical Rectification from 〈110〉 Zinc-Blende Crystals*, Appl. Phys. Lett. **64**, 1324 (1994).
[9] Y. Sasaki, A. Yuri, K. Kawase, and H. Ito, *Terahertz-Wave Surface-Emitted Difference Frequency Generation in Slant-Stripe-Type Periodically Poled LiNbO3 Crystal*, Appl. Phys. Lett. **81**, 3323 (2002).
[10] C. D'Amico, A. Houard, M. Franco, B. Prade, A. Mysyrowicz, A. Couairon, and V. T. Tikhonchuk, *Conical Forward THz Emission from Femtosecond-Laser-Beam Filamentation in Air*, Phys. Rev. Lett. **98**, 1 (2007).
[11] T. I. Oh, Y. J. Yoo, Y. S. You, and K. Y. Kim, *Generation of Strong Terahertz Fields Exceeding 8 MV/Cm at 1kHz and Real-Time Beam Profiling*, Appl. Phys. Lett. **105**, 041103 (2014).
[12] J. Liu, J. Dai, S. L. Chin, and X. C. Zhang, *Broadband Terahertz Wave Remote Sensing Using Coherent Manipulation of Fluorescence from Asymmetrically Ionized Gases*, Nat. Photonics **4**, 627 (2010).
[13] T. I. Oh, Y. S. You, N. Jhajj, E. W. Rosenthal, H. M. Milchberg, and K. Y. Kim, *Intense Terahertz Generation in Two-Color Laser Filamentation: Energy Scaling with Terawatt Laser Systems*, New J. Phys. **15**, 1 (2013).
[14] M. C. Beard, G. M. Turner, and C. A. Schmuttenmaer, *Terahertz Spectroscopy*, J. Phys. Chem. B **106**, 7146 (2002).
[15] C. Jansen, S. Wietzke, O. Peters, M. Scheller, N. Vieweg, M. Salhi, N. Krumbholz, C. Jördens, T. Hochrein, and M. Koch, *Terahertz Imaging: Applications and Perspectives*, Appl. Opt. **49**, E48 (2010).
[16] B. Kaulakys, V. Gontis, and G. Vilutis, *Ionisation of Rydberg Atoms by Subpicosecond Electromagnetic Pulses*, Lith. J. Phys. **33**, 290 (1993).
[17] G. Segschneider, F. Jacob, T. Löffler, H. G. Roskos, S. Tautz, P. Kiesel, and G. Döhler, *Free-Carrier Dynamics in Low-Temperature-Grown GaAs at High Excitation Densities Investigated by Time-Domain Terahertz Spectroscopy*, Phys. Rev. B - Condens. Matter Mater. Phys. **65**, 1252051 (2002).
[18] T. J. Wang, S. Yuan, Y. Chen, J. F. Daigle, C. Marceau, F. Théberge, M. Châteauneuf, J. Dubois, and S. L. Chin, *Toward Remote High Energy Terahertz Generation*, Appl. Phys. Lett. **97**, 111108 (2010).
[19] N. Stojanovic and M. Drescher, *Accelerator- and Laser-Based Sources of High-Field Terahertz Pulses*, Journal of Physics B: Atomic, Molecular and Optical Physics.
[20] B. Clough, J. Dai, and X. C. Zhang, *Laser Air Photonics: Beyond the Terahertz Gap*, Mater. Today **15**, 50 (2012).
[21] G. Rodriguez and G. L. Dakovski, *Scaling Behavior of Ultrafast Two-Color Terahertz Generation in Plasma Gas Targets: Energy and Pressure Dependence*, Opt. Express **18**, 15130 (2010).
[22] J. Bernhardt, W. Liu, S. L. Chin, and R. Sauerbrey, *Pressure Independence of Intensity Clamping during Filamentation: Theory and Experiment*, Appl. Phys. B Lasers Opt. **91**, 45 (2008).
[23] I. Dey, K. Jana, V. Y. Fedorov, A. D. Koulouklidis, A. Mondal, M. Shaikh, D. Sarkar, A. D. Lad, S. Tzortzakis, A. Couairon, and G. R. Kumar, *Highly Efficient Broadband Terahertz Generation from Ultrashort Laser Filamentation in Liquids*, Nat. Commun. **8**, 1 (2017).
[24] Q. Jin, Y. E, K. Williams, J. Dai, and X.-C. Zhang, *Observation of Broadband Terahertz Wave Generation from Liquid Water*, Appl. Phys. Lett. **111**, 071103 (2017).
[25] Q. Jin, J. Dai, E. Yiwen, and X. C. Zhang, *Terahertz Wave Emission from a Liquid Water Film under the Excitation of Asymmetric Optical Fields*, Appl. Phys. Lett. **113**, 261101 (2018).
[26] Y. Chu, X. Liang, L. Yu, Y. Xu, L. Xu, L. Ma, X. Lu, Y. Liu, Y. Leng, R. Li, and Z. Xu, *High-Contrast 2.0 Petawatt Ti:Sapphire Laser System*, Opt. Express **21**, 29231 (2013).
[27] G. Q. Liao, Y. T. Li, C. Li, S. Mondal, H. A. Hafez, M. A. Fareed, T. Ozaki, W. M. Wang, Z. M. Sheng, and J. Zhang, *Terahertz Emission from Two-Plasmon-Decay Induced Transient Currents in Laser-Solid Interactions*, Phys. Plasmas **23**, 013104 (2016).
[28] A. Gopal, P. Singh, S. Herzer, A. Reinhard, A. Schmidt, U. Dillner, T. May, H.-G. Meyer, W. Ziegler, and G. G. Paulus, *Characterization of 700 MJ T Rays Generated during High-Power Laser Solid Interaction*, Opt. Lett. **38**, 4705 (2013).
[29] G. Liao, Y. Li, H. Liu, G. G. Scott, D. Neely, Y. Zhang, B. Zhu, Z. Zhang, C. Armstrong, E. Zemaityte, P. Bradford, P. G. Huggard, D. R. Rusby, P. McKenna, C. M. Brenner,

N. C. Woolsey, W. Wang, Z. Sheng, and J. Zhang, *Multimillijoule Coherent Terahertz Bursts from Picosecond Laser-Irradiated Metal Foils*, Proc. Natl. Acad. Sci. U. S. A. **116**, 3994 (2019).

[30] A. Gopal, S. Herzer, A. Schmidt, P. Singh, A. Reinhard, W. Ziegler, D. Brömmel, A. Karmakar, P. Gibbon, U. Dillner, T. May, H. G. Meyer, and G. G. Paulus, *Observation of Gigawatt-Class THz Pulses from a Compact Laser-Driven Particle Accelerator*, Phys. Rev. Lett. **111**, 074802 (2013).

[31] Y. S. You, T. I. Oh, and K. Y. Kim, *Off-Axis Phase-Matched Terahertz Emission from Two-Color Laser-Induced Plasma Filaments*, Phys. Rev. Lett. **109**, 183902 (2012).

[32] P. Prem Kiran, S. Bagchi, S. R. Krishnan, C. L. Arnold, G. R. Kumar, and A. Couairon, *Focal Dynamics of Multiple Filaments: Microscopic Imaging and Reconstruction*, Phys. Rev. A **82**, 013805 (2010).

[33] S. Edition, *Chapter I Theoretical Background*, Vol. 42 (1967).

[34] J. B. Bates, *Fourier Transform Spectroscopy*, Comput. Math. with Appl. **4**, 73 (1978).

[35] C. Kübler, R. Huber, and A. Leitenstorfer, *Ultrabroadband Terahertz Pulses: Generation and Field-Resolved Detection*, Semicond. Sci. Technol. (2005).

[36] C. Winnewisser, P. Uhd Jepsen, M. Schall, V. Schyja, and H. Helm, *Electro-Optic Detection of THz Radiation in LiTaO3, LiNbO3 and ZnTe*, Appl. Phys. Lett. **70**, 3069 (1997).

[37] and J. Y. Zhihui Lu, Dongwen Zhang, Chao Meng, Lin Sun, Zhaoyan Zhou, Zengxiu Zhao, *Polarization-Sensitive Air-Biased-Coherent-Detection for Terahertz Wave*, Appl. Phys. Lett. **101**, 081119 (2012).

[38] J. Dai, J. Liu, and X. C. Zhang, *Terahertz Wave Air Photonics: Terahertz Wave Generation and Detection with Laser-Induced Gas Plasma*, IEEE J. Sel. Top. Quantum Electron. **17**, 183 (2011).

[39] P. U. Jepsen, R. H. Jacobsen, and S. R. Keiding, *Generation and Detection of Terahertz Pulses from Biased Semiconductor Antennas*, J. Opt. Soc. Am. B **13**, 2424 (1996).

[40] I. Brener, D. Dykaar, A. Frommer, L. Pfeiffer, and K. West, *Terahertz Emission from Electric Field Singularities in Biased Semiconductors*, Conf. Proc. - Lasers Electro-Optics Soc. Annu. Meet. **21**, 83 (1996).

[41] A. D. Koulouklidis, V. Y. Fedorov, and S. Tzortzakis, *Spectral Bandwidth Scaling Laws and Reconstruction of THz Wave Packets Generated from Two-Color Laser Plasma Filaments*, Phys. Rev. A **93**, (2016).

[42] M. Li, W. Li, Y. Shi, P. Lu, H. Pan, and H. Zeng, *Verification of the Physical Mechanism of THz Generation by Dual-Color Ultrashort Laser Pulses*, Appl. Phys. Lett. **101**, (2012).

[43] K.-Y. Kim, J. H. Glownia, A. J. Taylor, and G. Rodriguez, *Terahertz Emission from Ultrafast Ionizing Air in Symmetry-Broken Laser Fields*, Opt. Express (2007).

[44] T. D. Arber, K. Bennett, C. S. Brady, A. Lawrence-Douglas, M. G. Ramsay, N. J. Sircombe, P. Gillies, R. G. Evans, H. Schmitz, A. R. Bell, and C. P. Ridgers, *Contemporary Particle-in-Cell Approach to Laser-Plasma Modelling*, Plasma Phys. Control. Fusion **57**, 1 (2015).

[45] C. K. Birdsall and A. B. Langdon, *Plasma Physics via Computer Simulation* (Taylor & Francis, 2005).

[46] J. M. Dawson, *Particle Simulation of Plasmas*, Rev. Mod. Phys. **55**, 403 (1983).

[47] L. Berge, S. Skupin, F. Lederer, G. Me, J. Yu, J. Kasparian, E. Salmon, J. P. Wolf, M. Rodriguez, and L. Wo, *Multiple Filamentation of Terawatt Laser Pulses in Air*, **92**, 225002 (2004).

[48] R. Beigang, S. G. Biedron, S. Dyjak, F. Ellrich, M. W. Haakestad, D. Hübsch, T. Kartaloglu, E. Ozbay, F. Ospald, N. Palka, U. Puc, E. Czerwinska, A. B. Sahin, A. Sešek, J. Trontelj, A. Švigelj, H. Altan, A. D. van Rheenen, and M. Walczakowski, *Comparison of Terahertz Technologies for Detection and Identification of Explosives*, Terahertz Physics, Devices, Syst. VIII Adv. Appl. Ind. Def. **9102**, 91020C (2014).

[49] M. R. Leahy-Hoppa, M. J. Fitch, and R. Osiander, *Terahertz Spectroscopy Techniques for Explosives Detection*, Anal. Bioanal. Chem. **395**, 247 (2009).

[50] J. F. Federici, B. Schulkin, F. Huang, D. Gary, R. Barat, F. Oliveira, and D. Zimdars, *THz Imaging and Sensing for Security Applications - Explosives, Weapons and Drugs*, Semicond. Sci. Technol. **20**, (2005).

[51] W. L. Chan, J. Deibel, and D. M. Mittleman, *Imaging with Terahertz Radiation*, Reports on Progress in Physics.

[52] R. Matsunaga and R. Shimano, *Nonequilibrium BCS State Dynamics Induced by Intense Terahertz Pulses in a Superconducting NbN Film*, Phys. Rev. Lett. **109**, 187002 (2012).

[53] M. Liu, H. Y. Hwang, H. Tao, A. C. Strikwerda, K. Fan, G. R. Keiser, A. J. Sternbach, K. G. West, S. Kittiwatanakul, J. Lu, S. A. Wolf, F. G. Omenetto, X. Zhang, K. A. Nelson, and R. D. Averitt, *Terahertz-Field-Induced Insulator-to-Metal Transition in Vanadium Dioxide Metamaterial*, Nature **487**, 345 (2012).

[54] X. C. Zhang, A. Shkurinov, and Y. Zhang, *Extreme Terahertz Science*, Nat. Photonics **11**, 16 (2017).

[55] D. Nicoletti and A. Cavalleri, *Nonlinear Light–Matter Interaction at Terahertz Frequencies*, Adv. Opt. Photonics **8**, (2016).